\def\be{\begin{equation}}

\def\ee{\end{equation}}
\def\ba{\begin{array}}
\def\ea{\end{array}}

\documentclass[12pt]{article}
\usepackage{amssymb}
\begin{document}

\baselineskip=20pt
\setcounter{page}{1} \centerline{\LARGE\bf Invariants for a Class of Nongeneric} \vspace{3ex}
\setcounter{page}{1} \centerline{\LARGE\bf Three-qubit States} \vspace{4ex}

\begin{center}

Bao-Zhi Sun$^a$, ~Shao-Ming Fei$^{a,b}$

\vspace{2ex}
\begin{minipage}{5in}

{\small $~^{a}$ Department of Mathematics, Capital Normal University, Beijing 100037}

{\small $~^{b}$ Max-Planck-Institute for Mathematics in the Sciences, 04103 Leipzig}

\end{minipage}
\end{center}

\vskip 1 true cm
\parindent=18pt
\parskip=6pt
\begin{center}
\begin{minipage}{5in}
\vspace{3ex} \centerline{\large Abstract} \vspace{4ex}

We investigate the equivalence of quantum states under local
unitary transformations. A complete set of invariants under local
unitary transformations is presented for a class of non-generic
three-qubit mixed states.
It is shown that two such states in this class are locally equivalent
if and only if all these invariants have equal values for them.

\bigskip
\medskip
\bigskip
\medskip

\vfill

\smallskip

\vfill
\end{minipage}
\end{center}

\bigskip
\bigskip

Quantum entangled states are playing very important roles in
quantum information processing and quantum
computation \cite{nielsen}. As the properties of entanglement of a
multipartite quantum system remain invariant under local unitary
transformations on the subsystems, the entanglement can be
characterized by all the invariants of the local unitary
transformations. The trace norms of realigned or partial transposed density
matrices in entanglement measure and separability criteria
are some of these invariants \cite{norm}. Invariants are also relevant in the discussions of
Bell inequalities \cite{MPR200,PR210} and teleportation,
etc. \cite{MPR222}. A complete set of invariants
gives rise to the classification of the
quantum states under local unitary transformations. Two quantum
states are locally equivalent if and only if all these invariants
have equal values for these states. For bipartite mixed states,
a generally non-operational method has
been presented to compute all the invariants of local unitary
transformations in \cite{Rains,Grassl}.
In \cite{Makhlin}, the invariants for general two-qubit systems are
studied and a complete set of 18 polynomial invariants is
presented. In \cite{Albeverio03,goswami}, complete sets of invariants for
some classes of density matrices have been presented.
The invariants for tripartite pure states have been also studied \cite{wl}.

In \cite{Linden99}, the invariants for three qubits states have been
discussed. A complete set of invariants for generic mixed states
are presented. In this paper we investigate the invariants for non-generic
three-qubit states, and present some new invariants
for a class of such nongeneric three-qubit states.
These invariants plus the invariants given in
\cite{Linden99} make the complete set of invariants for these
states. Any two of these density matrices are
locally equivalent if and only if all these invariants have
equal values for these density matrices.

The density matrices of three qubits  may be written as :
\begin{eqnarray}
\rho&=&\frac{1}{8}(I_2\otimes I_2\otimes
I_2+\alpha_i\sigma_i\otimes I_2\otimes I_2+\beta_i
I_2\otimes\sigma_i\otimes I_2+\gamma_iI_2\otimes I_2\otimes\sigma_i\nonumber\\
&&+R_{ij}\sigma_i\otimes\sigma_j\otimes I_2 +S_{ij}\sigma_i\otimes
I_2\otimes\sigma_j +T_{ij}I_2\otimes\sigma_i\otimes\sigma_j
+Q_{ijk}\sigma_i\otimes\sigma_j\otimes\sigma_k),
\end{eqnarray}
where $\sigma_i$ are pauli matrices, the repeated indices are assumed to
be summed over from $1$ to $3$. If one considers the qubits
as spin-$1/2$ particles, the three dimensional vectors $\alpha=(\alpha_1,\alpha_2,\alpha_3)$,
$\beta=(\beta_1,\beta_2,\beta_3)$, $\gamma=(\gamma_1,\gamma_2,\gamma_3)$
are their average spins in the state $\rho$, while
$R_{ij},\ S_{ij},\ T_{ij}$ are spin-spin correlators respectively,
and $Q_{ijk}$ are three spin correlators. Any local unitary
transformations on $\rho$ can be represented by three corresponding $3\times 3$
orthogonal real matrices of ``spin rotation", $L,\ M,\ N\in
SO(3,I\!\!R)$. The operation, $L\otimes M\otimes N$, transforms
$\rho$ according to the following rules: $\alpha\rightarrow
L\alpha,\ \beta\rightarrow M\beta,\ \gamma\rightarrow N\gamma,\
R\rightarrow LRM^T,\ S\rightarrow LSN^T,\ T\rightarrow MTN^T,\
Q^{(1)}\rightarrow LQ^{(1)}M^T\otimes N^T$, where the last relation
can be also written as: $Q^{(2)}\rightarrow MQ^{(2)}L^T\otimes N^T$ or
$Q^{(3)}\rightarrow NQ^{(3)}L^T\otimes M^T$, with
$$Q^{(1)}=\left(\matrix{
Q_{111}&Q_{112}&Q_{113}&Q_{121}&Q_{122}&Q_{123}&Q_{131}&Q_{132}&Q_{133}\cr
Q_{211}&Q_{212}&Q_{213}&Q_{221}&Q_{222}&Q_{223}&Q_{231}&Q_{232}&Q_{233}\cr
Q_{311}&Q_{312}&Q_{313}&Q_{321}&Q_{322}&Q_{323}&Q_{331}&Q_{332}&Q_{333}\cr
}\right),$$
$$Q^{(2)}=\left(\matrix{
Q_{111}&Q_{112}&Q_{113}&Q_{211}&Q_{212}&Q_{213}&Q_{311}&Q_{312}&Q_{313}\cr
Q_{121}&Q_{122}&Q_{123}&Q_{221}&Q_{222}&Q_{323}&Q_{231}&Q_{322}&Q_{323}\cr
Q_{131}&Q_{132}&Q_{133}&Q_{231}&Q_{232}&Q_{233}&Q_{331}&Q_{332}&Q_{333}\cr
}\right),$$
$$Q^{(3)}=\left(\matrix{
Q_{111}&Q_{121}&Q_{131}&Q_{211}&Q_{221}&Q_{231}&Q_{311}&Q_{321}&Q_{331}\cr
Q_{112}&Q_{122}&Q_{132}&Q_{212}&Q_{222}&Q_{232}&Q_{312}&Q_{322}&Q_{332}\cr
Q_{113}&Q_{123}&Q_{133}&Q_{213}&Q_{223}&Q_{233}&Q_{313}&Q_{323}&Q_{333}\cr
}\right).$$

In \cite{Linden99}, the authors discussed the local equivalence for
states of three spin-$1/2$ particles by fixing a canonical
point on a generic orbit. And they presented an explicit finite
set of polynomial invariants to characterize the generic equivalent classes
of three spin-$1/2$ particles under local unitary transformations.
In this paper we discuss the nongeneric classes in terms of
the canonical point.

We use the notations in  \cite{Linden99} by defining
$X_{ii'}=\sum_{jk}Q_{ijk}Q_{i'jk}$,
$Y_{jj'}=\sum_{ik}Q_{ijk}Q_{ij'k}$,
$Z_{kk'}=\sum_{ij}Q_{ijk}Q_{ijk'}$. The matrices $X,\ Y,\ Z$ are
hermitian and positive. So they can be diagonalized by proper
rotations $L,\ M,\ N$ respectively. And one can always arrange
these diagonal entries in decreasing order. In generic case, these
entries are different, and the only remaining transformations
which keep the $X,Y$ and $Z$ in these diagonal forms are the local
unitary transformations which induce the orthogonal
transformations such that $L,M$ and $N$ are one of the matrices
$\mbox{diag}(1,-1,-1)$, $\mbox{diag}(-1,1,-1)$, or
$\mbox{diag}(-1,-1,1)$.

On the generic orbits $X,\ Y$ and $Z$ have different
eigenvalues respectively, and all the components of $
L\alpha,\ M\beta,\ N\gamma$ are
not zero when $X,\ Y,\ Z$ are transformed into diagonal matrices
by some $L$, $M$ and $N$. A canonical point on the generic orbit is uniquely given by
specifying that $X,\ Y$ and $Z$ are diagonalized by some $L$, $M$ and $N$
and have distinct eigenvalues, while all
the components of $L\alpha$ ($M\beta$ resp. $N\gamma$) have the same sign.
The generic orbits are thus parameterized by the components of $\alpha,\ \beta,\ \gamma,\
X,\ Y,\ Z$ and $Q$ at the canonical point on an orbit.
The invariants that characterize the generic orbits are \cite{Linden99}
\begin{eqnarray}&&trX^r,trY^r,trZ^r,\ \
\alpha^TX^{r-1}\alpha,\beta^TY^{r-1}\beta,\gamma^TZ^{r-1}\gamma\\
&&(\alpha,\ X\alpha,\ X^2\alpha),\ (\beta,\ Y\beta,\ Y^2\beta),\
(\gamma,\ Z\gamma,\ Z^2\gamma),\\
&&\alpha^TX^{r-1}RY^{s-1}\beta,\ \alpha^TX^{r-1}SZ^{s-1}\gamma,\
\beta^TY^{r-1}TZ^{s-1}\gamma,\\
&&\sum_{ijk}(X^{r-1}\alpha)_i(Y^{s-1}\beta)_j(Z^{t-1}\gamma)_kQ_{ijk}
=\alpha^TX^{r-1}Q^{(1)}(Y^{s-1}\otimes
Z^{t-1})(\beta\otimes\gamma),\end{eqnarray}
where $r,s,t=1,2,3$, $(\alpha,\ X\alpha,\ X^2\alpha)=
\epsilon_{ijk}\alpha_i(X\alpha)_j(X^2\alpha)_k$. From these invariants,
one can uniquely solve the parameters $\alpha$, $\beta$, $\gamma$, $R$, $S$, $T$
and $Q$. Hence the equivalent class of generic states are completely described by these invariants.

For the nongeneric case, the invariants above are not enough to
specify an orbit under local unitary transformations.
We first discuss the non generic case that there
is one and only one zero component in $\alpha$ or $\beta$ or $\gamma$
at the canonical point on an orbit.
Without loss of generality, we suppose $\alpha_3=0$ and
$\alpha_i\beta_j\gamma_k\neq0,\ i=1,2,\ j,k=1,2,3$. Here we may
 specify a canonical point by specifying that
 $\alpha_1\geq0,\ \alpha_2\geq0$ and the
components of $\beta$ (resp. $\gamma$) have the same sign.
From the invariants (2)-(5)
one can solve $\alpha_i,\beta_j,\gamma_k$ and all the components of $R,S,T,Q$,
except for $R_{3i},S_{3i}$ and $Q_{3ij}$, $i,j=1,2,3$. To solve these
components, we need the following extra invariants:
\begin{equation}\label{6}
(\alpha,\ X\alpha,\ RY^{r-1}\beta),\ (\alpha,\ X\alpha,\
SZ^{r-1}\gamma),\ (\alpha,\ X\alpha,\ Q^{(1)}(Y^{r-1}\otimes
Z^{s-1})(\beta\otimes\gamma)),
\end{equation}
where $r,s=1,2,3$. In
the case $\alpha_3=0$, there are three invariants from the first and
the second expressions of (\ref{6}) respectively:
$$\alpha_1\alpha_2(x_2^2-x_1^2)\sum
R_{3i}y_i^{2(r-1)}\beta_i,~~~~\alpha_1\alpha_2(x_2^2-x_1^2)\sum
S_{3i}z_i^{2(r-1)}\gamma_i,$$
and nine invariants from the last formula:
$$\alpha_1\alpha_2(x_2^2-x_1^2)\sum
Q_{3ij}y_i^{2(r-1)}z_j^{2(s-1)}\beta_i\gamma_j.$$
They can be written in matrix forms:
$$
\alpha_1\alpha_2(x_2^2-x_1^2)\Lambda FR_3,~~~
\alpha_1\alpha_2(x_2^2-x_1^2)\Theta GS_3,~~~
\alpha_1\alpha_2(x_2^2-x_1^2)(\Lambda F)\otimes (\Theta G)Q_3,
$$
where:
$$
\Lambda=\left(\matrix{ 1&1&1\cr
y_1^2&y_2^2&y_3^2\cr y_1^4&y_2^4&y_3^4}\right),\
\Theta=\left(\matrix{ 1&1&1\cr z_1^2&z_2^2&z_3^2\cr
z_1^4&z_2^4&z_3^4}\right),
$$
$F=\mbox{diag}(\beta_1,\beta_2,\beta_3)$, $G=\mbox{diag}(\gamma_1,\gamma_2,\gamma_3)$,
$R_3=(R_{31},R_{32},R_{33})^T$, $S_{3}=(S_{31},S_{32},S_{33})^T$,
$Q_3=(Q_{311},Q_{312},Q_{313},Q_{321},Q_{322},Q_{323},Q_{331},Q_{332},Q_{333})^T$.
Since we assume that $\mbox{det}(\Lambda F)$ and
$\mbox{det}(\Theta G)$ are not zero, the components
$R_{3i},\ S_{3i}$ and $Q_{3ij}$ can be solved from these new invariants.

When the zero entry is $\alpha_1$ or $\alpha_2$, the corresponding components
$R_{1i},\ S_{1i},\ Q_{1ij}$ or $R_{2i},\
S_{2i},\ Q_{2ij}$ can be also computed respectively by using the invariants in (\ref{6}).

Similarly, when the zero entry is in $\beta$ or $\gamma$, the
extra invariants we need are
\begin{equation}(\beta,\ Y\beta,\ R^TX^{r-1}\alpha),\
(\beta,\ Y\beta,\ TZ^{r-1}\gamma),\ (\beta,\ Y\beta,\
Q^{(2)}(X^{r-1}\otimes Z^{s-1})(\alpha\otimes\gamma))
\end{equation} or
\begin{equation}(\gamma,\ Z\gamma,\ S^TX^{r-1}\alpha),\
(\gamma,\ Z\gamma,\ T^TY^{r-1}\beta),\ (\gamma,\ Z\gamma,\
Q^{(3)}(X^{r-1}\otimes Y^{s-1})(\alpha\otimes\beta))
\end{equation} respectively, where $r,s=1,2,3$.

As an examples of the above non generic states we set $R_{ij}=S_{ij}=T_{ij}=0$,
$\alpha_1=\alpha_2=\beta_1=\beta_2=\gamma_1=\gamma_2=Q_{111}$,
$\gamma_3=\beta_3=Q_{333}$, $Q_{ijk}=0$ except for $i=j=k=1,2,3$ and $Q_{111}\neq
Q_{222}\neq Q_{333}$.
According to the relations \cite{0302176} between entries of
$\rho$ and the real parameters related to $\rho$,
we have the density matrix with real parameters:
\begin{equation}
\rho(a,b,c)=\frac{1}{8}\left(\matrix{
1+3c&ax&ax&0&ax&0&0&a+bi\cr a\bar{x}&1-c&0&ax&0&ax&a-bi&0\cr
a\bar{x}&0&1-c&ax&0&a-bi&ax&0\cr
0&a\bar{x}&a\bar{x}&1-c&a+bi&0&0&ax\cr
a\bar{x}&0&0&a-bi&1+c&ax&ax&0\cr
0&a\bar{x}&a+bi&0&a\bar{x}&1+c&0&ax\cr
0&a+bi&a\bar{x}&0&a\bar{x}&0&1+c&ax\cr
a-bi&0&0&a\bar{x}&0&a\bar{x}&a\bar{x}&1-3c\cr
}\right),
\end{equation}
where $a=Q_{111},b=Q_{222},c=Q_{333},x=1-i, \bar{x}=1+i$.
When $a=\pm0.1,b=0,c\in (-0.3,0.3)$, the matrix $\rho(a,b,c)$ is positive.
Moreover for the same $c$, the invariants
given above have the same values for $\rho(0.1,0,c)$ and
$\rho(-0.1,0,c)$. So $\rho(0.1,0,c)$ and
$\rho(-0.1,0,c)$ are equivalent under local unitary transformations,
i.e. they belong to the same orbit.

When there are two zero entries in the components (on the canonical point) of
$\alpha,\beta,\gamma$, the problem is much more complicate. We first
consider the case that the two zero entries are in
different vectors.  We suppose that $\alpha_3=\beta_3=0$ and specify the canonical
point in the orbit by $\alpha_i\geq0$, $\beta_i\geq0$, $i=1,2$. Then
the invariants given above can determine all the components of
$R,S,T,Q$ except for $R_{33}$ and $Q_{33i}$, $i=1,2,3$. By introducing
new invariants  $trRR^T$ and $trQ^{(3)T}Z^{r-1}Q^{(3)}$ we can solve
$R_{33}^2$ and $Q_{33i}^2$ in terms of these invariants. And the new invariants
$(Q^{(3)T}Z^{r-1}\gamma)^2$, $r=1,2,3$ can determine the values of
$Q_{33i}Q_{33j},\ i\neq j$. Therefore in this case we can get the value of $|R_{33}|$ and
can determine $Q_{33i}$ up to a same sign from the invariants.

For the case that the two zeros are in a same vector, we
suppose $\alpha_2=\alpha_3=0$, then we can specify the
canonical point as $\alpha_1\geq0$ and $\beta_i$ (resp.
$\gamma_i$) have same sign. In this case the invariants given by
(2)-(8) can only give us the values of $R_{1i},S_{1i},T_{ij}$ and
$Q_{1ij}$, where $i,j=1,2,3$. The invariants $trRY^{r-1}R^TX^{s-1},\
trSZ^{r-1}S^TX^{s-1},$ and $trQ^{(1)T}Q^{(1)}(Y^{s-1}\otimes
Z^{t-1})$ can determine the values of $R_{2i}^2$, $\ R_{3i}^2$, $S_{2i}^2$, $\ S_{3i}^2$
and $Q_{kij}^2$, $k=2,3$, $i,j=1,2,3$. By using the invariants
$(X^{r-1}RY^{s-1}\beta)^2$, $(X^{r-1}SZ^{s-1}\gamma)^2$ and
$((Y^{r-1}\otimes Z^{s-1})Q^{(1)T}X^{t-1}\alpha)^2$ we can further fix the values
of $R_{ki}R_{kj}$, $S_{ki}S_{kj}$ and
$Q_{kil}Q_{kjl},Q_{kli}Q_{klj}$, where $k=2,3$, $i,j,l=1,2,3$,
$i\neq j$. So we can determine the values of the remaining
components up to some sign.

When the zero entries are located in other places of the components
of $\alpha$, $\beta$ and $\gamma$, the new invariants can be similarly obtained.
We summarize these invariants as follows:
\begin{eqnarray}&&trRY^{r-1}R^TX^{s-1},\ trSZ^{r-1}S^TX^{s-1},\
trTZ^{r-1}T^TY^{s-1},\\
&&trQ^{(1)T}X^{r-1}Q^{(1)}(Y^{s-1}\otimes Z^{t-1}),\\
&&(X^{r-1}RY^{s-1}\beta)^2,\ (Y^{r-1}R^TX^{s-1}\alpha)^2,\
(X^{r-1}SZ^{-1}\gamma)^2,\nonumber\\
&&(Z^{r-1}S^TX^{s-1}\alpha)^2,\
(Y^{r-1}TZ^{s-1}\gamma)^2,\ (Z^{r-1}T^TY^{s-1}\beta)^2,\\
&&((Y^{r-1}\otimes Z^{s-1})Q^{(1)T}X^{t-1}\alpha)^2,\
((X^{r-1}\otimes Z^{s-1})Q^{(2)T}Y^{t-1}\beta)^2,\nonumber\\
&&((X^{r-1}\otimes Y^{s-1})Q^{(3)T}Z^{t-1}\gamma)^2.
\end{eqnarray}

The sign of the parameters can be further fixed for some detailed cases.
For example, with respect to the case that the two zeros are in different
vectors, if not all of the
invariants of $(\beta,\ Y\beta,\ TZ^{r-1}\gamma)$ or $(\alpha,\
X\alpha,\ SZ^{r-1}\gamma)$ are zero, then the entries of
$T_{3i}$ or $S_{3i}$ are not all zero. We can then use invariants
$(\alpha,\ X\alpha,\ RTZ^{r-1}\gamma)$ or $(\beta,\ Y\beta,\
R^TSZ^{r-1}\gamma)$ to determine the sign of $R_{33}$ . And use the invariants
$(\alpha,\ X\alpha,\ Q^{(1)}(Y^{r-1}\otimes Z^{s-1})\tilde{T})$ or
$(\beta,\ Y\beta,\ Q^{(2)}(X^{r-1}\otimes Z^{s-1})\tilde{S})$ to
determine the sign of $Q_{33i}$. Where
$\tilde{T}=(T_{11},T_{12},T_{13},T_{21},T_{22},T_{23},T_{31},T_{32},T_{33})^T$
and similarly for $\tilde{S}$.

We have investigated the invariants for non-generic three-qubit
states, and presented a complete set of invariants
for a class of  nongeneric three-qubit quantum mixed states.
Any two of these density matrices are
locally equivalent if and only if all these invariants have
equal values for these density matrices. For more complicated cases,
e.g. there are more zero entries in the components of $\alpha,\
\beta,\ \gamma$ in the canonical point of an orbit, by adding
some new invariants we can also
determine the parameters $R,\,S,\,T,\,Q$ up to a sign.
However we only know how to determine the sign by adding more new invariants
for some special cases. The problem for general cases is still
open.

\end{document}